\documentclass[preprint,12pt]{elsarticle}
\usepackage[utf8]{inputenc}
\usepackage{cmap}
\usepackage{amssymb}
\usepackage{amsmath}
\usepackage{amsthm}

\usepackage{graphicx}
\usepackage{doi}
\usepackage{lineno}
\usepackage{rotating}
\usepackage{multirow}
\usepackage{xcolor}
\usepackage{scalerel}

\journal{Pattern Recognition}

\begin{document}

\begin{frontmatter}

\title{Knowledge-driven Subword Grammar Modeling for Automatic Speech Recognition in Tamil and Kannada}

\author{Madhavaraj A\corref{cor1}\fnref{label1}}
\ead{madhavaraja@iisc.ac.in}
\author{Bharathi Pilar\fnref{label2}}
\ead{bharathi.pilar@gmail.com }
\author{Ramakrishnan A. G.\corref{cor1}\fnref{label1}}
\ead{agr@iisc.ac.in}
\fntext[label2]{Electrical Engineering, Indian Institute of Science\\
Bangalore, Karnataka, India}
\fntext[label2]{University College Mangalore, Karnartaka, India}
\cortext[cor1]{Corresponding Authors}

\begin{abstract}

In this paper, we present specially designed automatic speech recognition (ASR) systems for the highly agglutinative and inflective languages of Tamil and Kannada that can recognize unlimited vocabulary of words. We use subwords as the basic lexical units for recognition and construct subword grammar weighted finite state transducer (SG-WFST) graphs for word segmentation that captures most of the complex word formation rules of the languages. We have identified the following category of words (i) verbs, (ii) nouns, (ii) pronouns, and (iv) numbers. The prefix, infix and suffix lists of subwords are created for each of these categories and are used to design the SG-WFST graphs. We also present a heuristic segmentation algorithm that can even segment exceptional words that do not follow the rules encapsulated in the SG-WFST graph. Most of the data-driven subword dictionary creation algorithms are computation driven, and hence do not guarantee morpheme-like units and so we have used the linguistic knowledge of the languages and manually created the subword dictionaries and the graphs. Finally, we train a deep neural network acoustic model and combine it with the pronunciation lexicon of the subword dictionary and the SG-WFST graph to build the subword-ASR systems. Since the subword-ASR produces subword sequences as output for a given test speech, we post-process its output to get the final word sequence, so that the actual number of words that can be recognized is much higher. Upon experimenting the subword-ASR system with the IISc-MILE Tamil and Kannada ASR corpora, we observe an absolute word error rate reduction of 12.39\% and 13.56\% over the baseline word-based ASR systems for Tamil and Kannada, respectively.

\end{abstract}


\end{frontmatter}


\section{Introduction}
Research in automated speech recognition (ASR) has led to significant advancements and finds many  applications in the field of industry and home automation, mobile telephony, healthcare, defence, etc. Most of the ASR research methodologies and techniques in the literature focusing mainly on English and other European languages, and no significant progress has been made for Indian languages, especially Dravidian languages such as Tamil and Kannada. One of the main challenges in developing a automatic speech recognition systems for these languages is the unavailability of standard speech and text corpora, and the other reason being its complicated morphological structure. This paper focuses on overcoming these limitations by building a reasonably good, ASR systems for both Tamil and Kannada languages using subword modeling.

Since the word-based ASR approach is impractical for these languages due to the ever increasing vocabulary size, it is not possible to sufficiently cover the large vocabulary of words without utilizing subword units \cite{bhref11}. The subword modelling technique, where each word is split into two or more morpheme-like units, decreases the out of vocabulary (OOV) rate; besides the smaller subword vocabulary reduces the model complexity significantly. It also makes it possible to build n-gram language models to cover millions of words~\cite{bh_ref10}. It still requires special solutions for the state-of-the-art neural network language models \cite{bhref12} to train an output layer that has a vast number of possible output symbols. DNNs trained on several thousand hours of speech have reduced the word error rate (WER) significantly compared to the traditional methods and achieve word-level accuracy of nearly 90\% for vocabulary sizes of about 200,000 for English ~\cite{hinton2012deep}. Such ASR systems are now widely used for commercial and entertainment purposes.

Various methods have been used for segmenting the word into its constituent subwords. One of the popular techniques is the morphological analyzer that is used for segmenting words into morphs and also to obtain other information such as stems and part-of-speech. The morphological analyzers have been used in works \cite{bhref13},\cite{bhref14}. Syllables and other rule-based systems have been used in \cite{bhref15},~\cite{laureys2002hybrid}. Data-driven algorithms have been used in \cite{hacioglu2003lexicon},~\cite{arisoy2007language}. In this paper, we employ knowledge-driven approach and use the morphological rules for word formation in constructing the subword grammar WFST which is then used in the decoding graph of our proposed subword-based ASR.

The rest of the paper is organised a follows. Section \ref{sec_2} gives an overall outline of the proposed subword modeling based ASR approach. Section \ref{sec_3} describes the subword dictionary construction procedure and the design of the subword grammar WFST for word segmentation. Further, a heuristics-based algorithm is also presented to segment exception words. In section \ref{sec_4}, we elaborate the procedure to create the final subword grammar WFST and subword lexicon WFST to be used in our proposed subword-based ASR. Also, the post-processing steps to convert the recognized subword sequence from the subword-based ASR back to word sequence is also discussed. Section \ref{sec_5} presents the experimental setup and performance of the proposed approach, followed by conclusion and future scope of this work.

\section{The proposed subword-based ASR approach}
\label{sec_2}
\begin{figure*}[!htbp]
\includegraphics[width=0.75\textwidth,height=0.75\textheight]{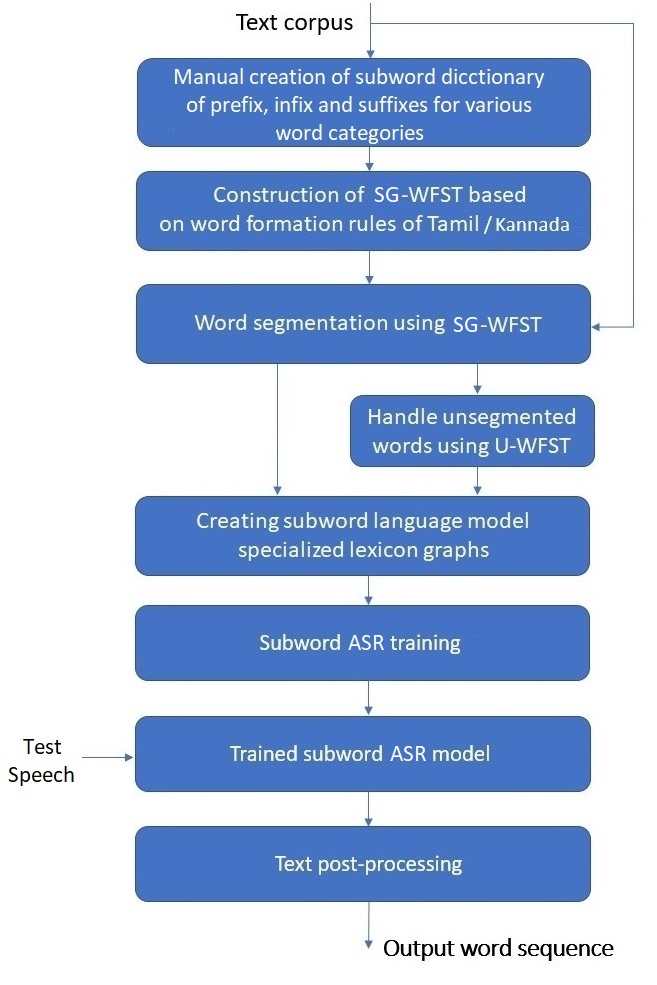}
\centering
\caption{Block diagram of the proposed subword modeling approach for building Tamil/Kannada ASR.}
	\label{fig_3_man}
\end{figure*}

The block diagram illustrating the training and testing of the proposed subword-based ASR is shown in figure \ref{fig_3_man}. First, we get the vocabulary of words from the input text corpus and create the list of prefixes, infixes and suffixes and categorize them into different word forms to create SG-WFST based on the word formation rules of the languages. Based on linguistic knowledge, we have formed the following category of words: (i) verbs, (ii) pronouns, (iii) numbers and (iv) nouns. These prefix, infix, and suffix lists of subwords are created such that a simple string concatenation of a prefix with one or more infixes and suffixes will form a valid Tamil word.

Next, we segment all the words in the text corpus using the SG-WFST graph, and the exception words that cannot be segmented is now segmented using another specially designed U-WFST graph. We then create the subword language model WFST and subword lexicon WFST and combine them with the trained DNN-based acoustic model to create the final decoding graph for the subword-ASR system. This decoding graph is used during testing to decode any given test speech. Since the subword-ASR produces a sequence of subwords, we post-process its output using deterministic rules as explained in \cite{sub_context} to get the final output word sequence. Since both Tamil and Kannada languages share almost similar word formation rules, the  word categories remains same for both the languages. 

For our Kannada experiments, we have used our \textit{IISc-MILE Kannada ASR Corpus}, which has 347 hours (275 hours for training and 72 hours for testing) of data. For the Tamil experiments, we use \textit{IISc-MILE Tamil ASR Corpus} and Microsoft Tamil data \cite{microsoftdata}. The combined Tamil dataset contains 217 hours (152 hours for training and 65 hours for testing) of transcribed speech data. The performance of the proposed subword-based ASR is compared with that of the baseline word-based ASR system. We have followed the procedure illustrated in \cite{asr_tamil} to build the baseline system. The \textit{IISc-MILE speech corpora} have now been made publicly available on OpenSLR \cite{mile_Kannada_asr_data,  mile_Tamil_asr_data}.

\section{Subword grammar WFST creation and segmentation}
\label{sec_3}

\subsection{WFST implementation of word segmentation using manually created subword dictionary}
\label{sec_3_1}

Initially, we have classified the words from the text corpus into 5 categories, namely, (i) past tense verbs, (ii) present and future tense verbs, (iii) nouns, (iv) pronouns and (v) numbers. This categorization is based on the linguistic knowledge of tense, number, person, gender, case, mood, and voice markers in Tamil and Kannada languages. We have analysed these five categories and divided each word into prefix, infix and suffix. The words are partitioned in such a way that the original word can be obtained by performing a simple string concatenation of the constituent subwords.

Word segmentation based on morphology involves the tedious task of constructing the subword grammar graphs which needs hard coding of a lot of grammar rules. However, the proposed approach of segmenting a word into its constituent prefix (modified form of root) and infix/suffixes (morphemes) has the following advantages: (i) It does not require hard coding of grammar rules during post-processing (ii) It does not involve modifications for lexicon preparation.

\begin{figure*}[!ht]
\includegraphics[width=\textwidth,height=0.26\textheight]{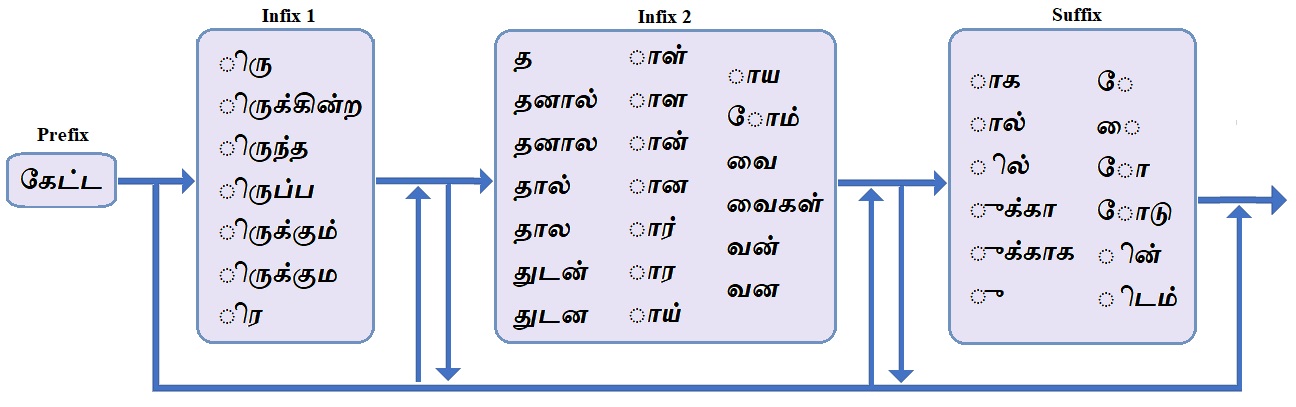}
\centering
\caption{Flowchart illustrating the rules to generate valid words that can be formed with an example past tense verb root \protect\scalerel*{\includegraphics{manual_tam_word_01.png}}{B} /keetta/  in Tamil}
	\label{fig_3_6}
\end{figure*}

\vspace{0.5cm}
\noindent
\textbf{Creating a subword list for words formed out of verb roots:}
In our framework, we have created two different sub-categories having distinct verb roots for the same verb for the words formed with (i) past tense markers and (ii) future and present tense markers. Since many other markers can be combined along with the tense markers to form valid words, we have carefully segregated the infixes and suffixes and sequentially ordered them to construct a flowchart such that there will be at least one valid path for most of the words that can be formed with the given root verb prefix. In figure \ref{fig_3_6}, we have shown the flowchart that forms most of the words having an example past tense verb root prefix (\scalerel*{\includegraphics{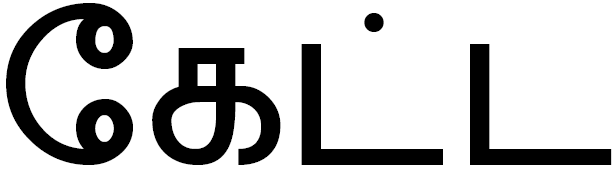}}{B}  /keetta/ , meaning ``asked'' in English). For simplicity, we have only given a handful of infix and suffix subwords in the flowchart. We see from this figure that we have taken care of the ordering of infixes that can be added to the verb root by cascading two sets of infixes and placing them between the prefix and suffix lists \cite{sub_tamverbs, sub_wordformtam02}. This configuration is particularly useful when building lexicon graph to be used in our subword ASR, since it reduces the space and time when performing graph search during decoding.

\begin{figure*}[!ht]
\includegraphics[width=\textwidth,height=0.26\textheight]{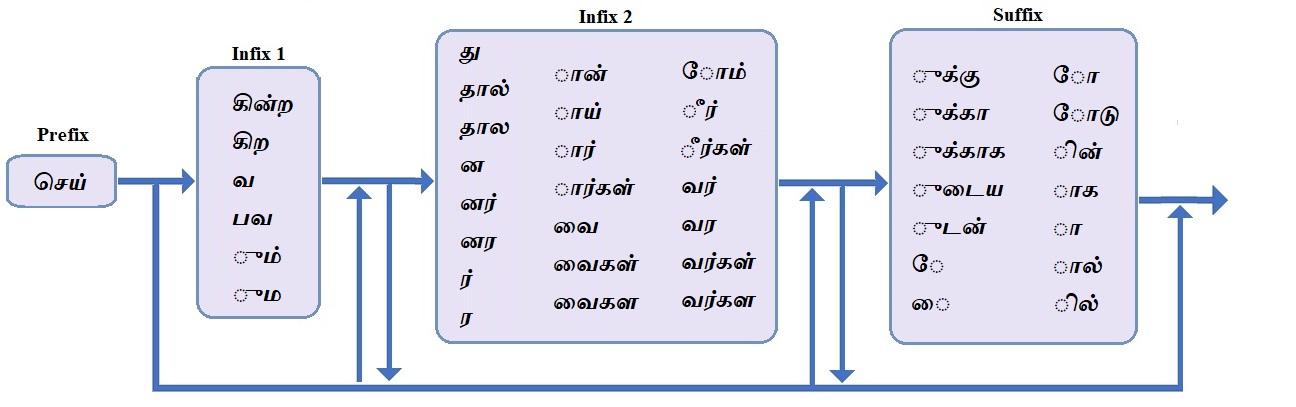}
\centering
\caption{Flowchart illustrating the rules to generate valid words that can be formed from a present/future tense verb root \protect\scalerel*{\includegraphics{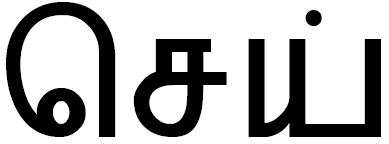}}{B} /sei/ in Tamil}
\label{fig_3_7}
\end{figure*}

Figure \ref{fig_3_7} shows the flowchart that forms present/future tense words that are derived from a single example verb root prefix (\scalerel*{\includegraphics{manual_tam_word_02.png}}{B} /sei/, meaning ``do'' in English). With these flowcharts shown in figures \ref{fig_3_6} and \ref{fig_3_7}, we can generate not only verbs but adjectives, nouns, infinitives, etc., that can be derived from the given set of verb root prefixes.

\begin{figure*}[!ht]
\includegraphics[width=\textwidth,height=0.29\textheight]{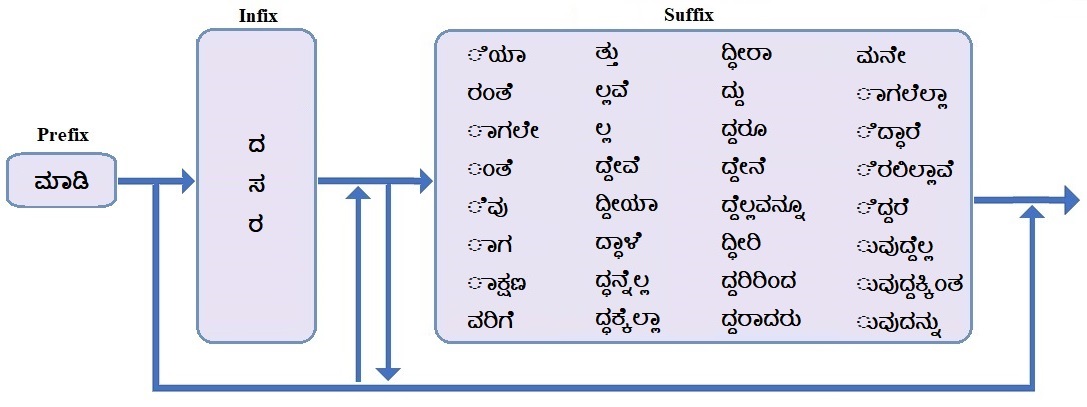}
\centering
\caption{Flowchart illustrating the rules to generate valid words from an example Kannada past tense verb root \protect\scalerel*{\includegraphics{manual_kan_word_03.png}}{B} /maadi/ , meaning 'to do' in English }
\label{fig_3_6_kan}
\end{figure*}

\begin{figure*}[!ht]
\includegraphics[width=\textwidth,height=0.29\textheight]{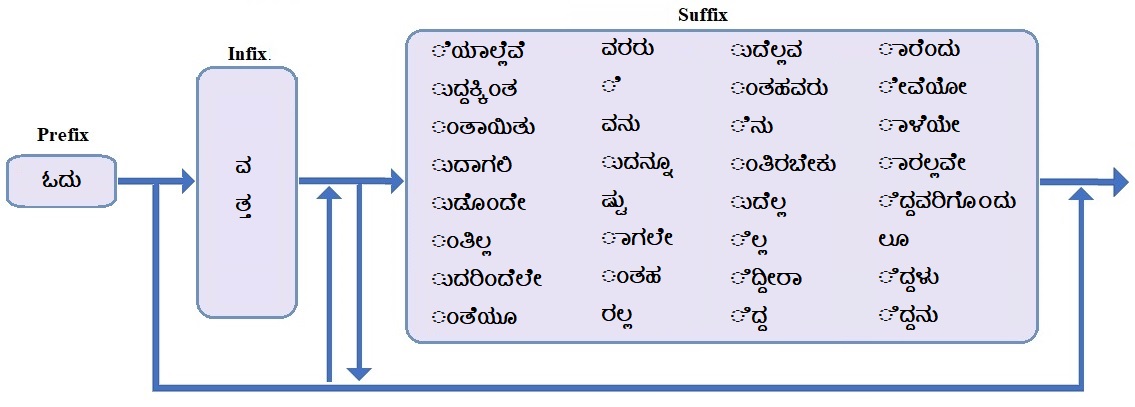}
\centering
\caption{Flowchart illustrating the rules to generate valid words from an example Kannada present/future tense verb root \protect\scalerel*{\includegraphics{manual_kan_word_04.png}}{B} /oodu/ (means 'to read').}
\label{fig_3_7_kan}
\end{figure*}

Similar flowcharts for Kannada language are shown. Figure \ref{fig_3_6_kan} shows the flowchart that forms most of the words having the past tense verb root prefix  ( \scalerel*{\includegraphics{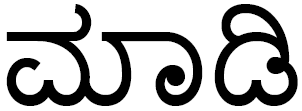}}{B} /maadi/, meaning ``do'' in English). In figure \ref{fig_3_7_kan}, we show the flowchart that forms present/future tense words that are derived from the example verb root prefix  (\scalerel*{\includegraphics{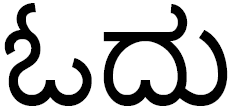}}{B} /oodu/, meaning ``read'' in English) \cite{prathibha2013development}. 

\begin{figure*}[!ht]
\includegraphics[width=0.8\textwidth,height=0.3\textheight]{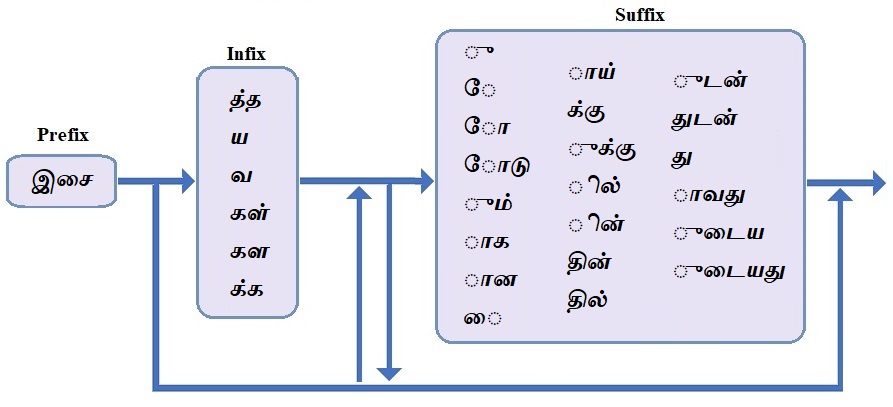}
\centering
\caption{Flowchart illustrating the rules to generate valid derived words from an example Tamil noun root word \protect\scalerel*{\includegraphics{manual_tam_word_05.png}}{B} /isai/}
\label{fig_3_8}
\end{figure*}

\vspace{0.5cm}
\noindent
\textbf{Creating subword list for words formed out of noun roots:} Similar to verbs, we have created a list of possible prefixes, infixes and suffixes for nouns. We have employed the morphological rules given in \cite{sub_tamnouns} for pruning the nouns to create the noun prefix list to be used in constructing the graph. The flowchart in figure \ref{fig_3_8} shows the Tamil words that can be formed from the noun (\scalerel*{\includegraphics{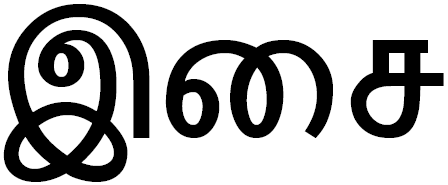}}{B} /isai/ , meaning ``music'' in English).

\begin{figure*}
\includegraphics[width=\textwidth,height=0.3\textheight]{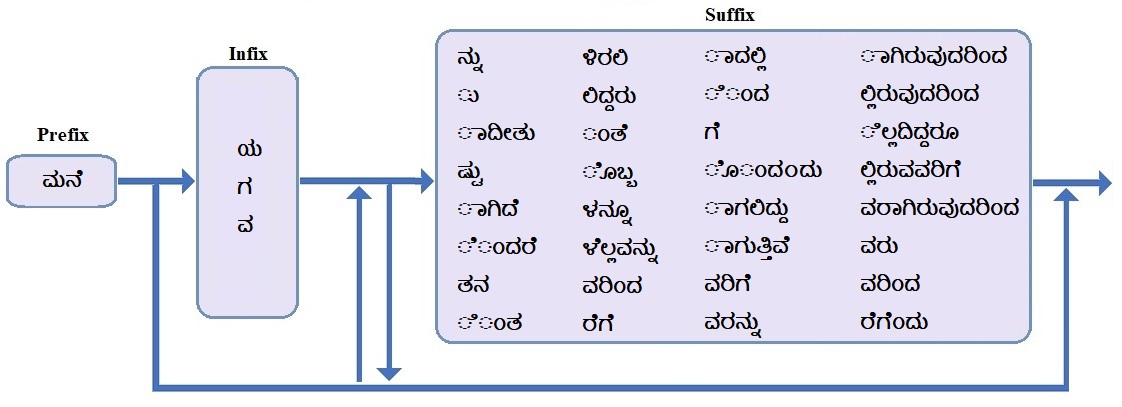}
\centering
\caption{Flowchart illustrating the rules to generate valid derived words from an example Kannada noun root word \protect\scalerel*{\includegraphics{manual_kan_word_06.png}}{B} /mane/}
\label{fig_3_8_kan}
\end{figure*}

A similar flowchart is shown for Kannada in figure \ref{fig_3_8_kan}, with the rules that form words from the noun \scalerel*{\includegraphics{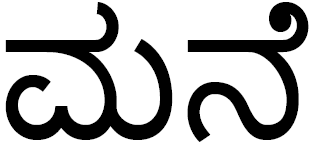}}{B}  /mane/, meaning ``house'' in English).

\begin{figure*}[!ht]
\includegraphics[width=0.8\textwidth,height=0.34\textheight]{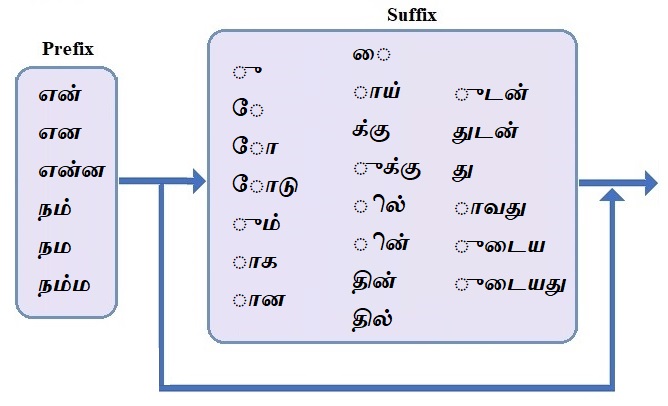}
\centering
\caption{Flowchart illustrating the rules to generate valid words that can be formed with the Tamil pronoun root words \protect\scalerel*{\includegraphics{manual_tam_word_07.png}}{B} /en/ and \protect\scalerel*{\includegraphics{manual_tam_word_08.png}}{B} /nam/. (Note that all possible prefixes for these roots have been added to the prefix list.)}
\label{fig_3_9}
\end{figure*}

\vspace{0.5cm}
\noindent
\textbf{Creating subword list for words formed out of pronoun roots:} The list of suffixes is almost the same for pronouns and nouns. Since no literature is available describing the rules for creating prefix list for pronouns, we have created a custom list of prefixes so that it can be incorporated in our framework. Figure \ref{fig_3_9} shows the flowchart for generating words that can be formed from the two pronoun roots \scalerel*{\includegraphics{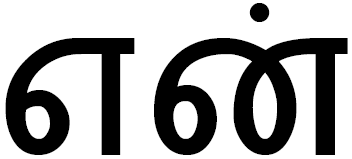}}{B} /en/ and \scalerel*{\includegraphics{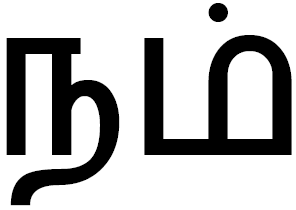}}{B} /nam/ (meaning ``my'' and ``our'' in English), respectively).

\begin{figure*}[!htbp]
\includegraphics[width=0.7\textwidth,height=0.34\textheight]{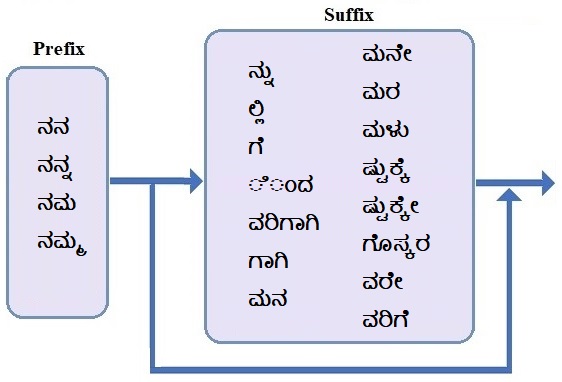}
\centering
\caption{Flowchart illustrating the rules to derive valid words for two Kannada pronoun root words \protect\scalerel*{\includegraphics{manual_kan_word_09.png}}{B} /nanna/ and  \protect\scalerel*{\includegraphics{manual_kan_word_10.png}}{B} /namma/.}
\label{fig_3_9_kan}
\end{figure*}

The same observation is true for Kannada. Figure \ref{fig_3_9_kan} shows the flowchart for deriving words from the two pronoun roots \scalerel*{\includegraphics{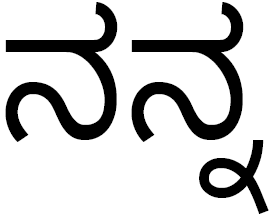}}{B}  /nanna/ and  \scalerel*{\includegraphics{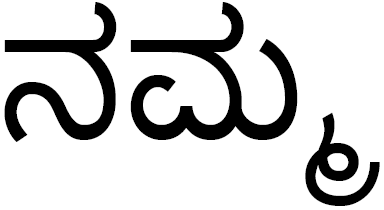}}{B} /namma/  (meaning ``my'' and ``our'' in English, respectively).

\vspace{0.5cm}
\noindent
\textbf{Creating subword list for words formed out of number roots:}
In the case of Tamil words formed from number root prefixes, in addition to inflexion, we need to handle agglutination. This has been performed by pruning some of the prefixes and including them in the infix list.

\begin{figure*}[!htbp]
\includegraphics[width=\textwidth,height=0.32\textheight]{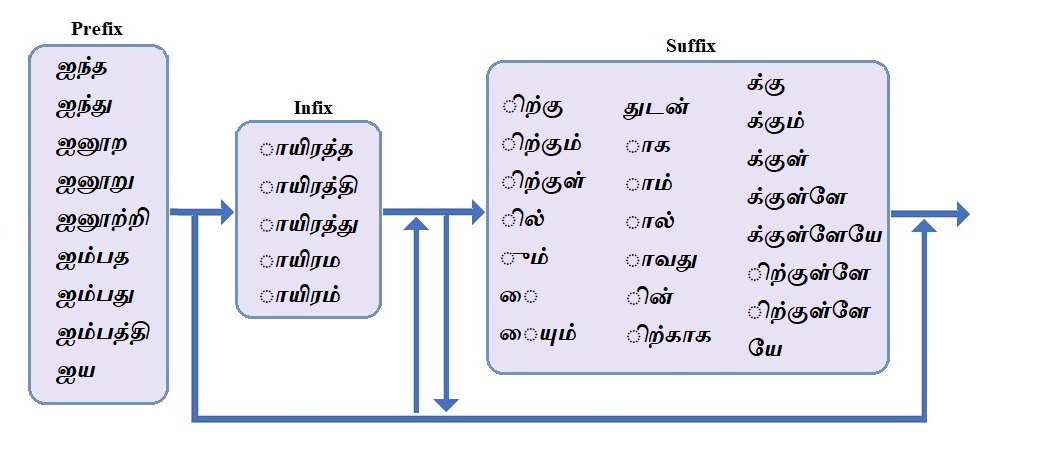}
\centering
\caption{An example flowchart illustrating the rules to generate valid words that can be formed from an example Tamil number root word \protect\scalerel*{\includegraphics{manual_tam_word_11.png}}{B} /aindhu/ (meaning five in English) with all possible prefixes added to the prefix list.}
\label{fig_3_10}
\end{figure*}

This pruning is performed based on our knowledge of the rules for numeric word formation. Figure \ref{fig_3_10} shows the flowchart that can generate all the words that can be formed from an example number word (\scalerel*{\includegraphics{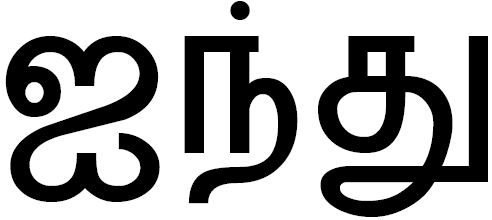}}{B} /aindhu/, meaning ``five'' in English).

\begin{figure*}
\includegraphics[width=\textwidth,height=0.28\textheight]{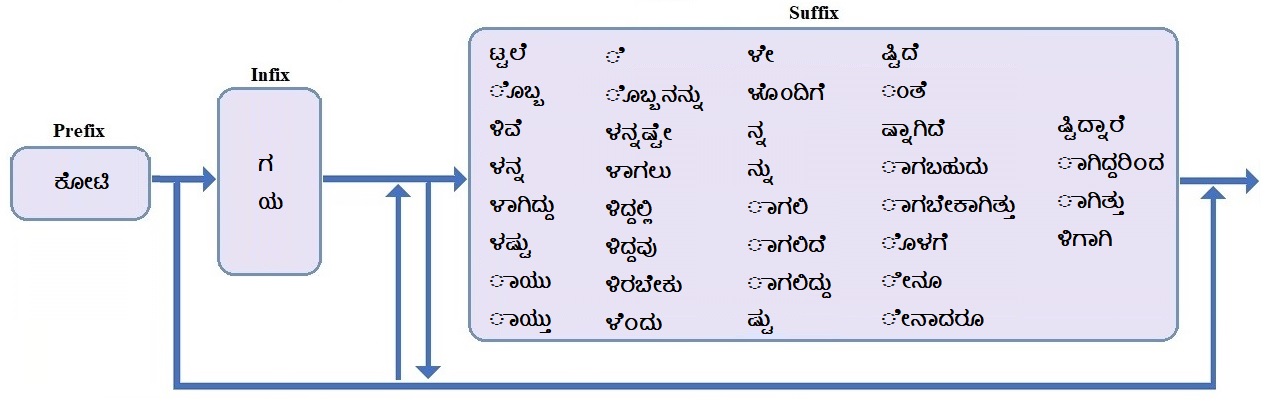}
\centering
\caption{Flowchart illustrating the rules in Kannada to generate valid words that can be formed from the number root \protect\scalerel*{\includegraphics{manual_kan_word_12.png}}{B} /koti/  with all possible suffixes added to the suffix list.}
\label{fig_3_10_kan}
\end{figure*}

Similarly we have created flowcharts for the numbers in Kannada. Figure \ref{fig_3_10_kan} shows the flowchart that can generate all the words that can be formed from an example number word   \scalerel*{\includegraphics{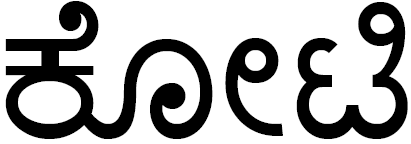}}{B} /koti/, meaning ``crore'' in English).

It is to be noted that, although we have used only five categories of words, other words belonging to adjectives, adverbs, infinitives, etc., categories can also be handled by our subword-ASR using the subword dictionary which we have manually created, if the lexicon WFST is constructed according to the manner explained as explained in section \ref{sec_4}.

\subsection{WFST implementation of word segmentation using manually created subword dictionary}
\label{sec_3_2}

\begin{figure*}[!ht]
\includegraphics[width=\textwidth,height=0.6\textheight]{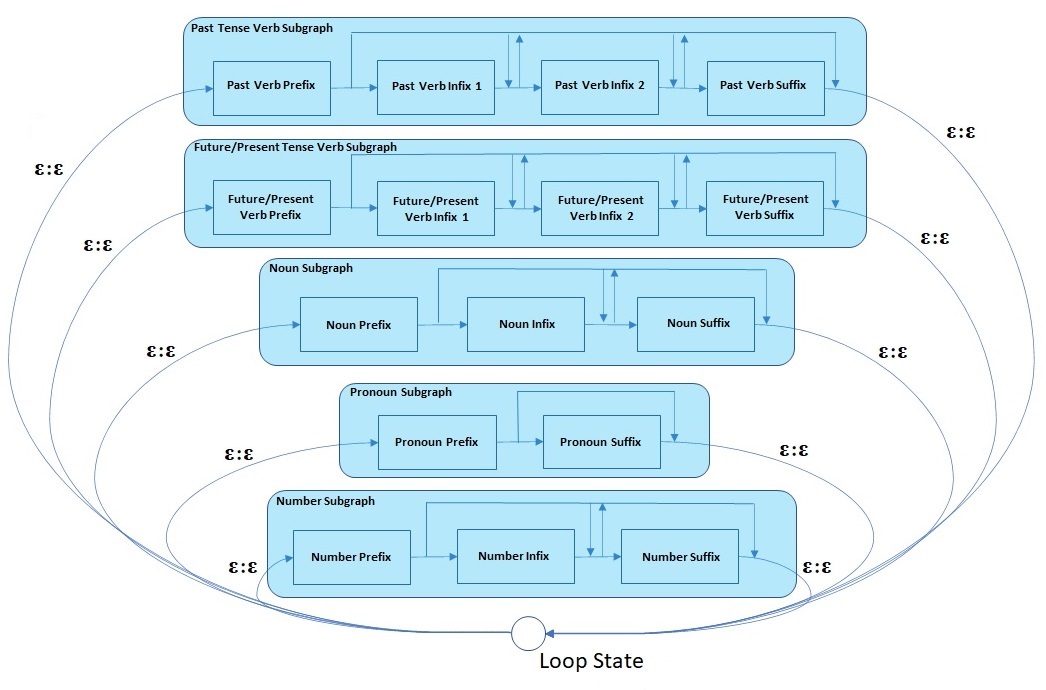}
\centering
\caption{Template of the manual subword weighted finite state transducer (SG-WFST) showing the subgraphs of the five categories connected to the loop state from both ends.}
	\label{fig_3_11}
\end{figure*}

The first step is to construct the subword grammar WFST (SG-WFST) using the flowcharts shown in figures \ref{fig_3_6}-\ref{fig_3_10_kan} as templates independently for each language. SG-WFST contains five subgraphs, one for each of the categories of words. (i.e., past tense verbs, present/future tense verbs, nouns, pronouns and numbers). This graph contains a single loop state to which the start and end nodes of the five subgraphs are connected through the transitions with $\epsilon$ labels at both input and output as shown in figure \ref{fig_3_11}. Each subgraph is a cascade of prefix, infix and suffix lists with appropriate skip transitions between them.

\begin{sidewaysfigure*}[!htbp]
\centering
\includegraphics[width=\textwidth]{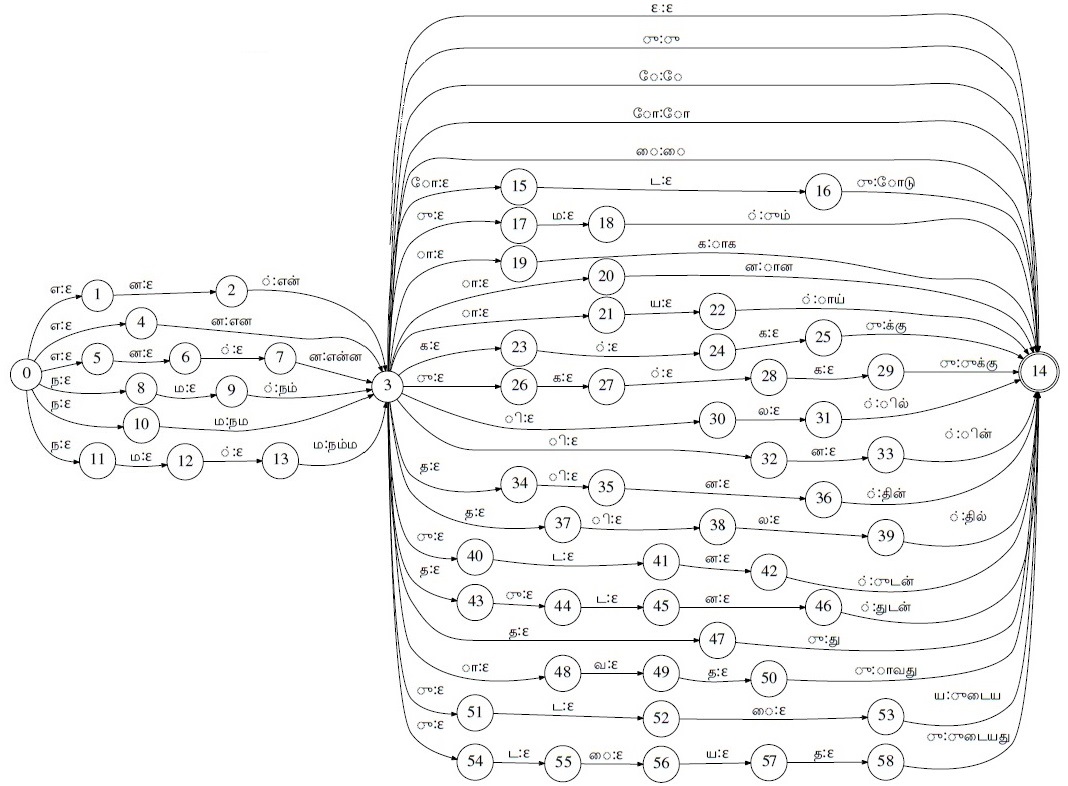}
\caption{The pronoun subgraph for Tamil which is used in the construction of SG-WFST. This graph is constructed using the flowchart shown in figure \ref{fig_3_9}.}
\label{fig_3_12M}
\end{sidewaysfigure*}

\begin{sidewaysfigure*}[!htbp]
\centering
\includegraphics[width=1.2\textwidth]{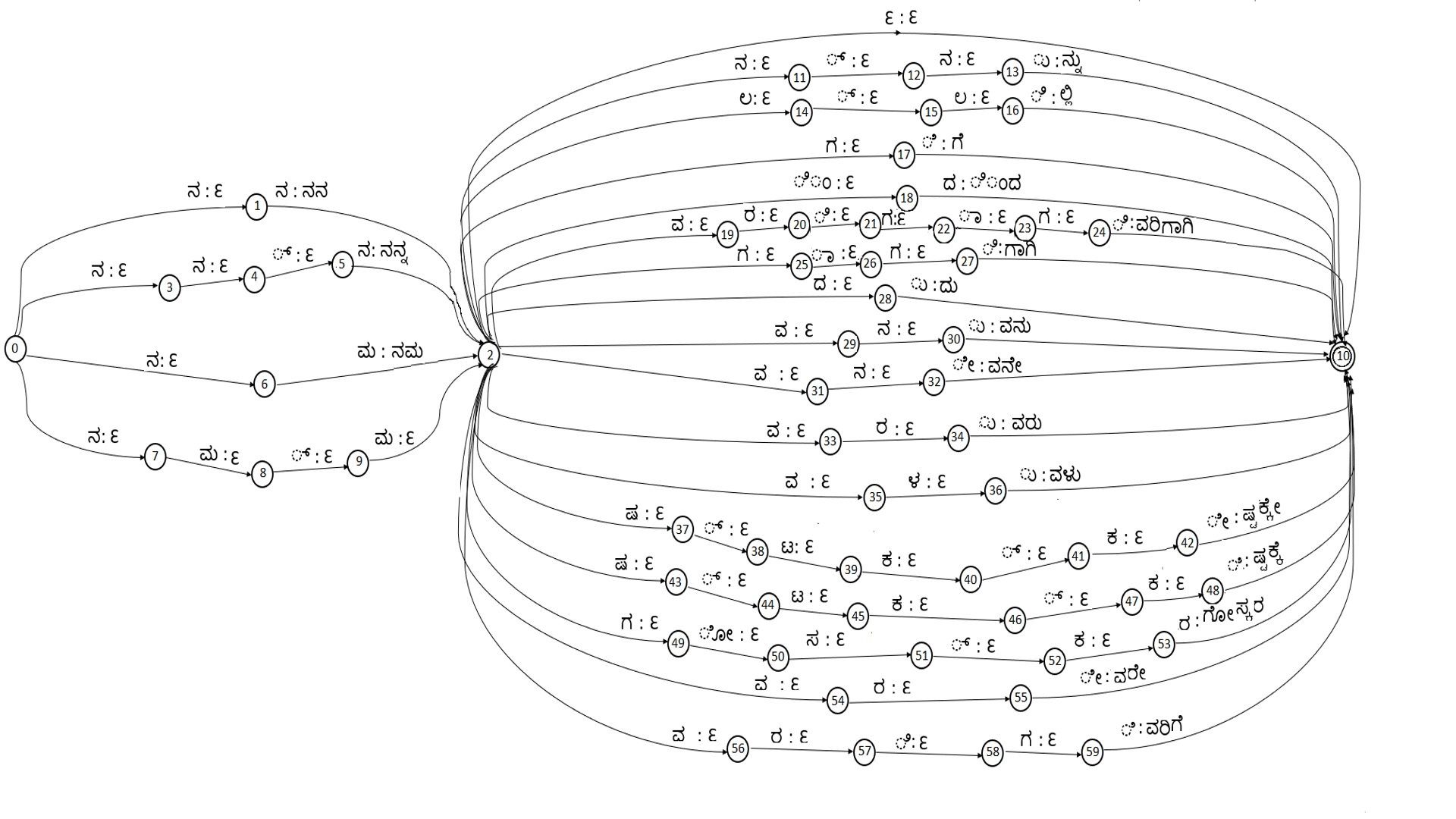}
\caption{The pronoun subgraph for Kannada which is used in the construction of SG-WFST. This graph is constructed using the flowchart shown in figure \ref{fig_3_9}.}
\label{fig_3_12}
\end{sidewaysfigure*}

Figure \ref{fig_3_12M} and  \ref{fig_3_12} shows an example WFST subgraph for pronoun category, for Tamil and Kannada respectively, which is a cascade of a set of smaller subgraphs of prefixes and suffixes. Every path from the start to the end state of the pronoun prefix subgraph contains a set of left-to-right transitions whose input labels are the characters in the subword that the path encodes and the output labels are the pronoun prefix subword string in the last transition and $\epsilon$ labels for the rest of the transitions. Similar structure is followed for pronoun suffix subgraphs as well, whose start state will be the end state of pronoun prefix subgraph, as shown in figure \ref{fig_3_12M} and \ref{fig_3_12}. Thus, we form SG-WFST by connecting the subgraphs of all the five categories in parallel.

\begin{figure*}[!ht]
\includegraphics[width=\textwidth,height=0.2\textheight]{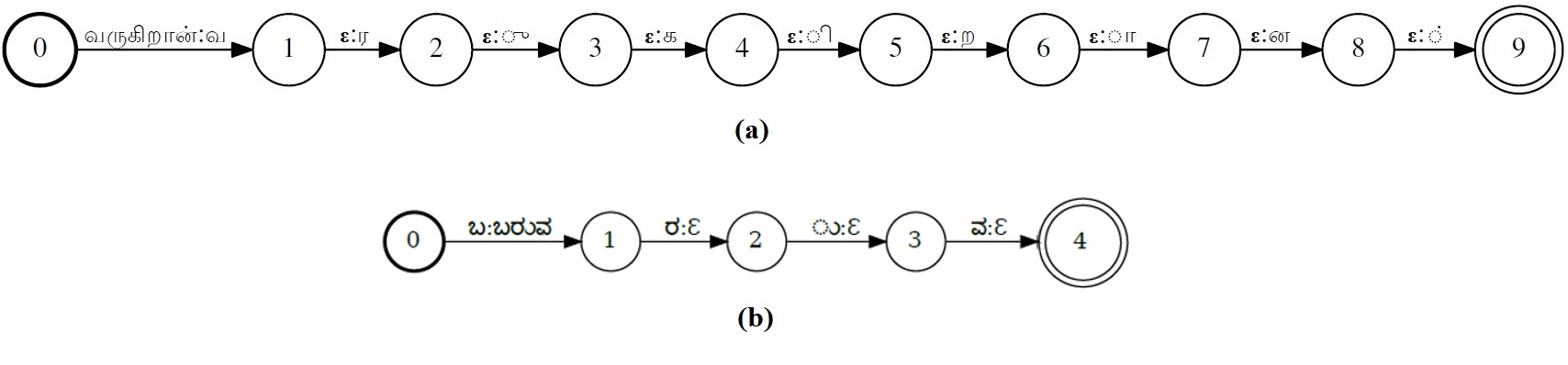}
\centering
\caption{W-WFST for an example (a) Tamil word  \protect\scalerel*{\includegraphics{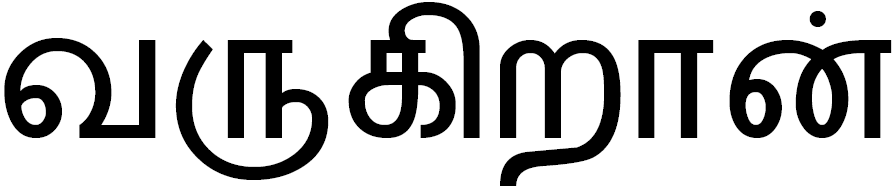}}{B} /varugiraan/ and (b) Kannada word \protect\scalerel*{\includegraphics{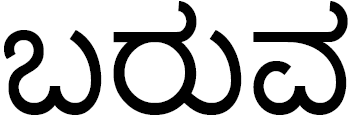}}{B}  /baruva/.}
	\label{fig_w_wfst}
\end{figure*}

In the second step, for each of the words in the vocabulary $\mathbb{V}$, we construct a W-WFST which contains only one path having unique start and end states. The input label for the first transition of the path is the word $w$ and $\epsilon$ for the rest of the transitions, while the output labels are the individual characters of $w$ as shown in figure \ref{fig_w_wfst} for an example Tamil and Kannada word.

Next, each of these W-WFSTs are composed with SG-WFST individually and the resulting WFST has only one path from the start to the end state. The output labels along this path provide the sequence of subwords segmented for the given word.

\subsection{Construction of U-WFST to handle unsegmented words}
\label{sec_3_3}
It is to be noted that not all the words in the vocabulary can be segmented using this method since SG-WFST has been manually created with a finite number of subwords. Due to agglutination and OOV issues, some of the words in the vocabulary do not follow the word formation rules encapsulated in the SG-WFST. However, we try our best to automatically segment such words with a new graph (named U-WFST) using some heuristic knowledge and assumptions.

\begin{figure*}[!ht]
\includegraphics[width=0.9\textwidth,height=0.32\textheight]{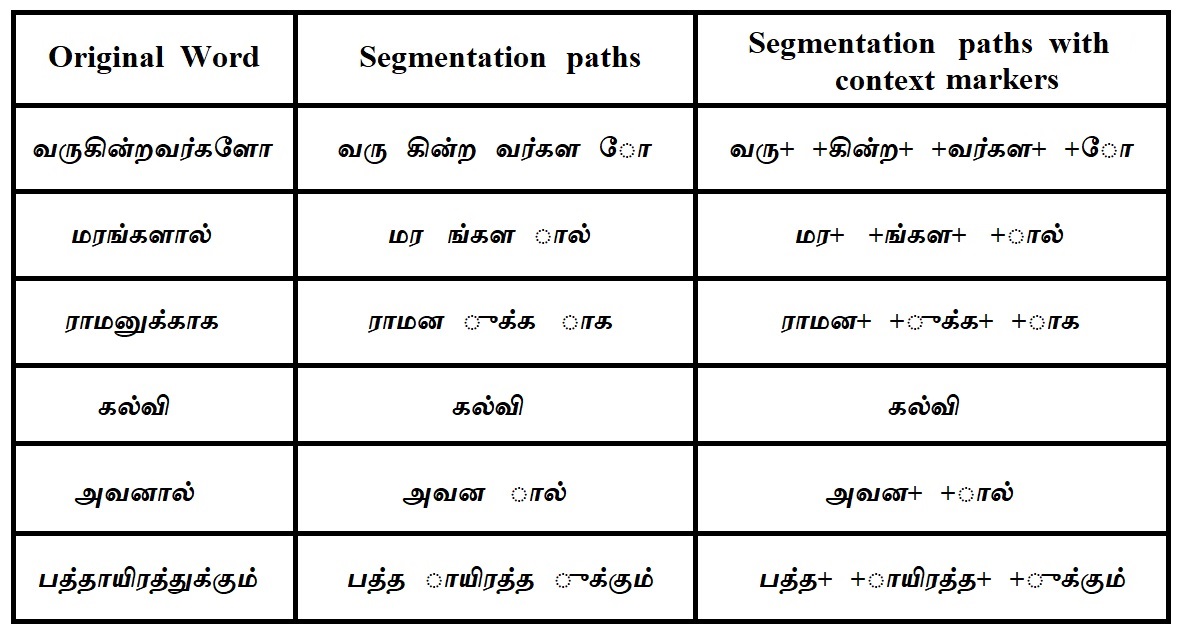}
\centering
\caption{Illustration of segmentation of some sample Tamil words and addition of context markers to the segmented subwords based on their category.}
	\label{fig_3_13M}
\end{figure*}

First, we we take the segmentation sequences of all the words in the vocabulary that can be segmented using the method explained in subsection \ref{sec_3_2}. Next, we add context identification markers (``+'') to every subword in the segments to identify whether they are prefixes, infixes, suffixes or singleton subwords \cite{sub_context}. Prefix subwords have markers only at the end of the string; suffix subwords have markers only at the beginning; infixes have markers at both the ends; whereas, singleton subwords have no markers. Figure \ref{fig_3_13M} show some example Tamil words, their segmented subword sequences and context markers added to the subwords.

Then we replace all the words in the text corpus by their context-marked segmentation sequences. Then, we create the dictionary of context-marked subwords (say $\mathbb{D}^+$) and calculate the unigram probability of their occurrences in the text corpus and create a list of 2-tuples $\{(s, \phi(s)) : \forall s \in \mathbb{D}^+ \}$. We then create another list of 2-tuples which contains all the individual characters of the language with a very small ($\delta$) weight associated with each of them. These two lists are merged to form the final list of 2-tuples (expressed in equation \ref{eqn_3_1}) to be used as labels and weights when constructing the U-WFST graph.

\begin{equation}
    \label{eqn_3_1}
    \mathbb{F} = \{(s, \phi(s)) : \forall s \in \mathbb{D}^+ \} \;\; \bigcup \;\; \{ (c, \delta) : \forall c \in \mathbb{C} \} 
\end{equation}
where $\mathbb{C}$ is the set of all valid characters of the language and the value of $\delta$ is chosen as 0.0001.

\begin{figure*}[!ht]
\includegraphics[width=\textwidth,height=0.72\textheight]{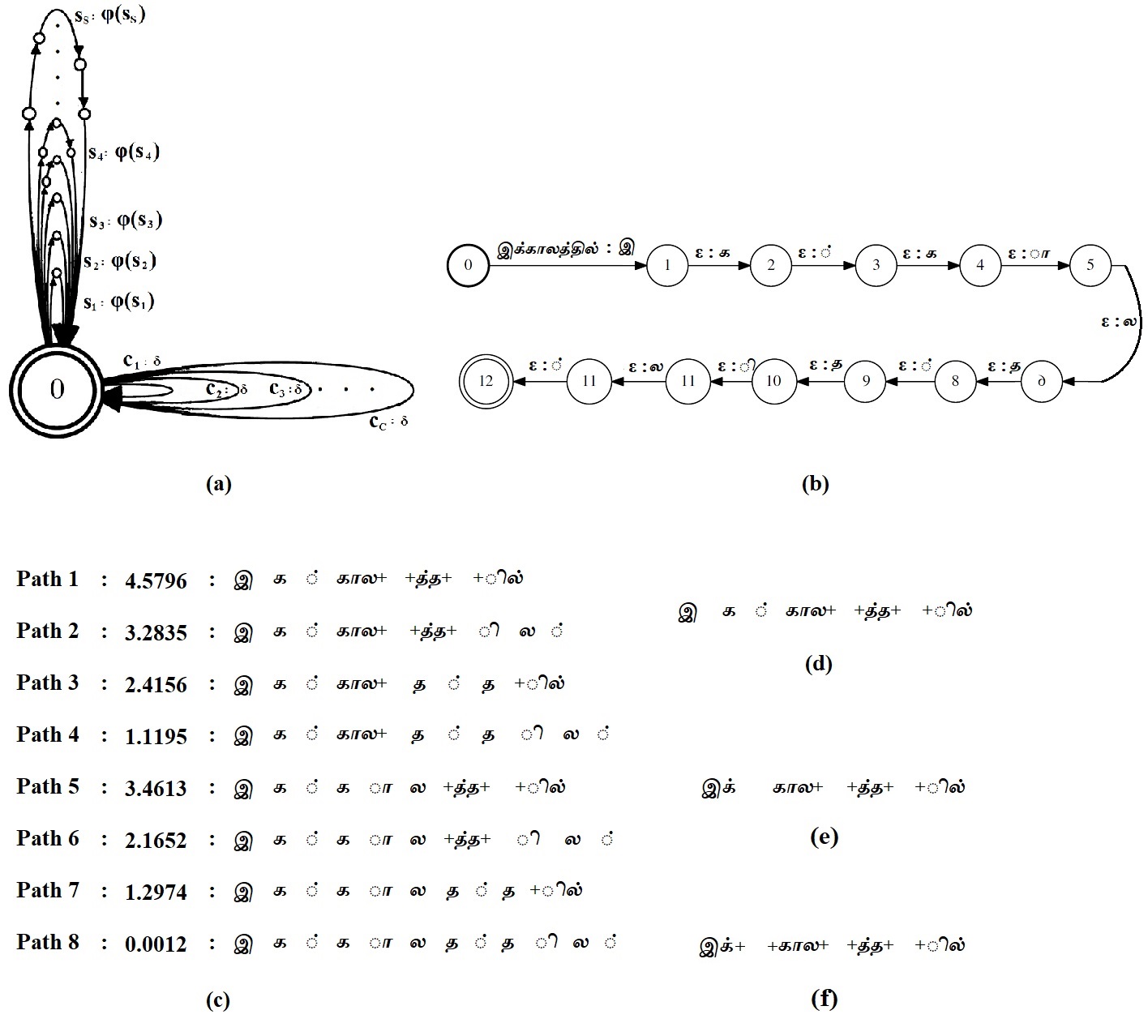}
\centering
\caption{Handling unsegmented words: (a) Template of U-WFST. (b) W-WFST for an example unsegmented word \protect\scalerel*{\includegraphics{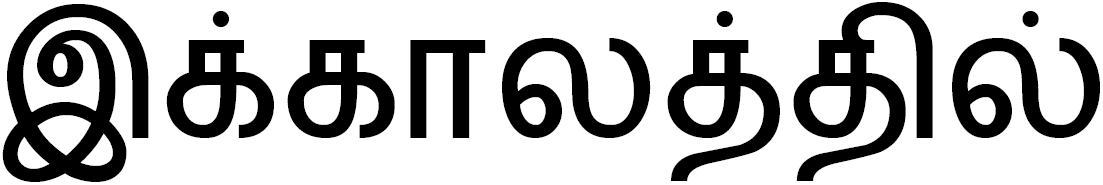}}{B} /ikkaalathil/. (c) possible segmentation paths obtained when composing (b) and (a). (d) optimal segmentation path chosen based on the proposed criteria. (e) segmentation obtained after merging contiguous 1-length subwords. (f) context markers added to the subword segments.}
\label{fig_3_14M}
\end{figure*}

Finally, we create the U-WFST by having one loop state and creating an unique path for each entry in the list $\mathbb{F}$ starting and ending in the loop state. The input labels for the transitions along a path are the individual characters of the subword, and the output labels are the subword string for the last transition and $\epsilon$ labels for the rest of the transitions. The weight of a path encoding the subword $s$ (or $c$) is set accordingly as $\phi(s)$ (or $\delta$) as shown in figure \ref{fig_3_14M}a.

\vspace{0.5cm}
\noindent\textbf{Segmentation using U-WFST and optimal subword sequence selection:} Next, we create W-WFSTs for all the unsegmented words and compose them with U-WFST. The resulting WFST obtained has many possible segmentation paths, out of which we choose the optimal one and prune the segments based on the following criteria.

\begin{itemize}
  \item \textbf{Criterion 1: } Merge contiguous segments if they are 1-character length subwords.
  
  \item \textbf{Criterion 2: } Consider only those paths where all the intermediate segments (whose individual lengths are more than 1) start and end with the context marker ``+''.
  
  \item \textbf{Criterion 3: } If there are many such paths satisfying criteria 1 and 2, the path which has the maximum weight is chosen.
  
  \item \textbf{Criterion 4: } For a given word, if there is no such path satisfying the above criteria, then we consider the whole word as one subword unit and add it to the subword dictionary (in the independent category).
\end{itemize}

We avoid the problem of over-segmentation of words by keeping a very low value for $\delta$ (so that the joint weight of the over-segmented subword sequence is very low) and the choice of the selection criteria. Figure \ref{fig_3_14M}b-f shows an example unsegmented word being segmented using U-WFST and how we select the best possible segmentation path. Thus, using both SG-WFST and U-WFST, we are able to segment any given word to subwords.

\section{Creation of n-gram subword language model and lexicon WFST for the proposed subword-ASR system}
\label{sec_4}
In this section, we describe the way to use subwords as vocabulary units and pre-process the original text corpus and learn n-gram subword language model (LM) and use it in our subword-ASR system. Since subword-ASR system uses subwords as vocabulary units, the first step is to segment the words in the transcription text and the LM text corpus into subwords using the segmentation technique explained in section \ref{sec_3}. We then add context-dependent markers to the segmented subwords based on their category. N-gram subword LM is then learnt from the segmented LM text corpus. In all our subword-ASR experiments, we have used $6^{th}$ order LM to learn the grammar of subword units. In normal ASRs, $3^{rd}$ order word-level LM is used. In our case, since each word is split into $2$ to $4$ subwords, only by increasing the subword LM order to $6$, we would match the performance and learn the grammar context of the $3^{rd}$ order word-level LM.

\begin{sidewaysfigure*}[!htbp]
\centering
\includegraphics[width=\textwidth]{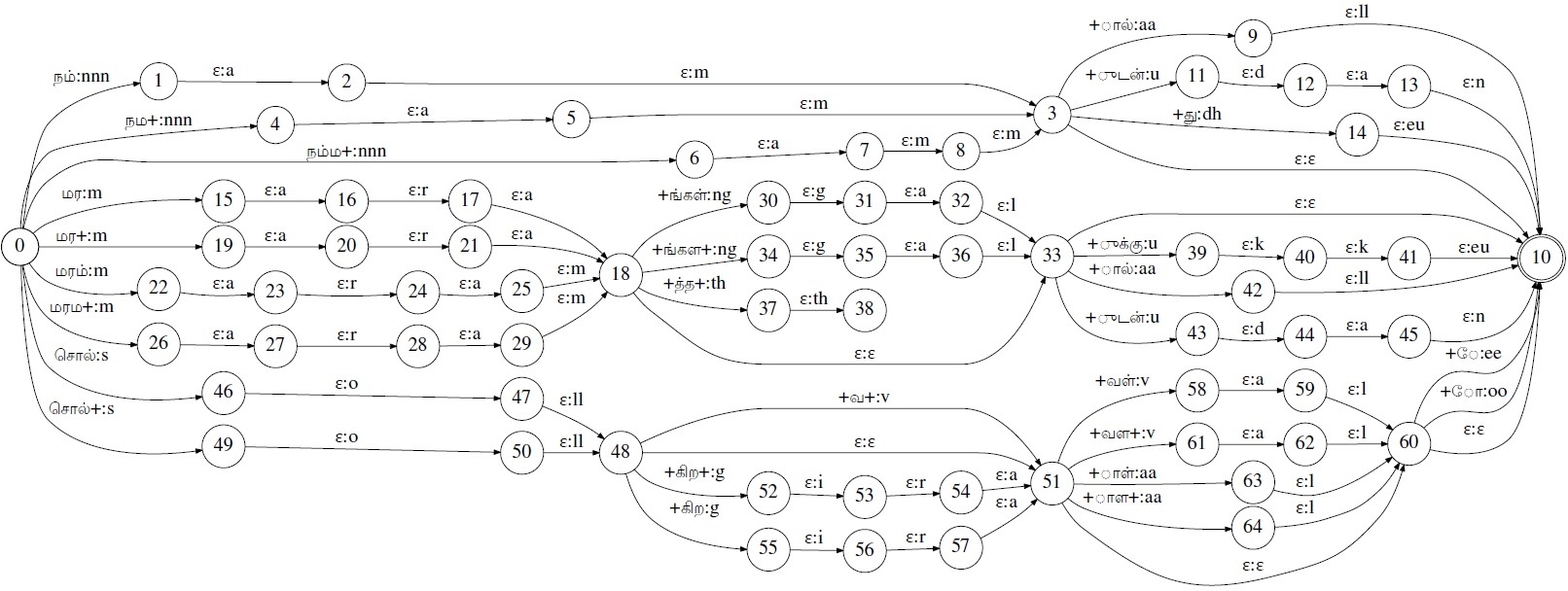}
\caption{Lexicon WFST created for an example context-marked subword dictionary $\mathbb{D^+}$ to be used in Tamil subword modeling ASR}
\label{fig_3_15M}
\end{sidewaysfigure*}

To create subword lexicon WFST, we first create the lists of prefixes, infixes, suffixes and singleton words in past-tense verb, future/present-tense verb, noun, pronoun, number and independent categories. Then, we construct the lexicon WFST using the same template as SG-WFST, except that the input labels are context-independent phones rather than characters. In addition, another subgraph for independent category is included. The pronunciation dictionary is created for every entry in the subword list using grapheme-to-phoneme (G2P) converters that we have built \cite{ramakrishnan2007grapheme}. Figure \ref{fig_3_15M} shows a sample lexicon WFST created for Tamil. It can be seen that by carefully designing the flow of transitions in the graph, we can encapsulate the word formation rules of the language so that a huge number of words can be recognized (after due post-processing) using this lexicon WFST.

Finally, we use the training set of the IISc-MILE ASR corpus to train the DNN acoustic model (AM) with the same procedure explained in \cite{asr_tamil}. This DNN AM is combined with the subword lexicon and language model WFSTs to create the decoding graph which we use for decoding the utterances from the test set.

The subword-ASR produces a sequence of subwords as output for a given utterance. Hence we need to post-process it to get a valid sequence of words. Post-processing is done by concatenating the subwords having the context markers (“+”) at either sides of the point of concatenation. Next, we remove all the context markers from the concatenated string to get the final post-processed word sequence which is then used to compute the WER. For the purpose of comparison, and to illustrate the benefits of subword-ASR for Tamil/Kannada, we have considered only the words in the training corpus to build the lexicon for the baseline word-based ASR system. Hence its OOV and WER on the test set are higher.

\section{Experimental setup and results}
\label{sec_5}
In this section we compare the performance of our proposed subword-ASR with respect to the baseline word-based ASR system for subword LMs of different orders. The performance of subword-based ASRs is compared with the baseline word-based ASRs in terms of OOV rate and WER to illustrate the significance of subword modeling to handle highly agglutinative languages like Tamil and Kannada.

The language models are learnt from a large text corpus of 4.4 million Tamil words and 8 million Kannada words. The size of dictionary of words that we have used in building our Tamil and Kannada baseline word-based ASR systems are 182771 and 201055, respectively. For, the subword-ASR systems, the Tamil and Kannada subword dictionaries that we have manually created contain 43856 and 58098 words, respectively.

\begin{table}[ht]
\centering 
  \caption{Comparison of the word error rates (WER) and out of vocabulary (OOV) rates of the subword-ASR system for Tamil language.}
  \begin{tabular}{| c | c |c| c| c |c| }
  \hline
    \multirow{2}{*}{Method} & \multicolumn{4}{c|}{Word Error Rate (\%)} & \multirow{2}{*}{OOV} \\[2pt]
	\cline{2-5}
	& 3-gram LM & 4-gram LM & 5-gram LM & 6-gram LM & \\[4pt] 
	\hline
	\hline
	Baseline & 24.70 & -NA- & -NA- & -NA- & 10.73 \\[4pt]
	\hline
	Manual method   & 18.04	& 15.93  & 	14.12  & 	 12.31 & 	1.68 \\
	\hline
\end{tabular}%
\label{tab_3_1}
\end{table}

\begin{table}[ht]
\centering 
  \caption{Comparison of the word error rates (WER) and out of vocabulary (OOV) rates of the subword-ASR system for Kannada language.}
  \begin{tabular}{| c | c |c| c| c |c| }
  \hline
    \multirow{2}{*}{Method} & \multicolumn{4}{c|}{Word Error Rate (\%)} & \multirow{2}{*}{OOV} \\ [2pt]
	\cline{2-5}
	& 3-gram LM & 4-gram LM & 5-gram LM & 6-gram LM & \\[4pt] 
	\hline
	\hline
	Baseline & 21.95 & -NA- & -NA- & -NA- & 8.64 \\[4pt]
	\hline
	Manual method   & 12.97	& 10.27 & 	9.47 & 	8.39 & 	1.12 \\
	\hline
\end{tabular}%
\label{tab_3_2}
\end{table}

Tables \ref{tab_3_1} and \ref{tab_3_2} compare the WER performance of the proposed subword ASR systems for subword LMs of different orders, with the baseline method. We have also listed the OOV rates to further justify the effectiveness of the subword modeling. OOV rate is defined as the ratio of number of words in the test corpus which are not present in the training corpus to the total number of words in the test corpus.

Since the subword dictionary creation and word segmentation methods have been carefully handcrafted to handle the agglutination and inflexion of Tamil and Kannada words, this method gives the best WER of 12.31\% which is an absolute improvement of 12.39\% over the baseline word-level ASR system in the case of Tamil, whereas best WER of 8.39\% (an absolute improvement of 13.56\% over the baseline word-level ASR system) is obtained for Kannada. However, the OOV rates of 1.68\%  and  1.12\% for Tamil and Kannada languages, respectively, arise due to proper nouns and other uncommon words present in the test set. Such words can also be easily handled by the ASR when they are included in the independent category without any additional manual effort.

\section{Conclusion and future work}
\label{sec_6}
In this paper, we have presented a novel design for Tamil and Kannada ASR systems based on subword modeling by manually creating the subword dictionary and using to design a specialized SG-WFST for word segmentation purpose. We have used linguistic knowledge and identified the following category of words (i) verbs, (ii) pronouns, (iii) numbers and (iv) nouns to create the prefix, infix and suffix lists to design the SG-WFST that capture the word formation rules of Tamil and Kannada. Experiments conducted on the IISc-MILE ASR corpus validate the significance of the proposed subword-ASR system in bringing down the OOV and WER by a huge margin.

The proposed SG-WFST creation and segmentation technique finds application not only in ASR but also in other areas of natural language processing. It can be used for applications like lemmatization, parts-of-speech tagging, text translation and transliteration, by simply changing the input and output labels of the graph. The same approach is also very useful for correction of the recognized text in the case of printed or handwritten text recognition systems, since they have the same issue of having to deal with unlimited word-vocabulary.



\begin{thebibliography}{20}
\expandafter\ifx\csname natexlab\endcsname\relax\def\natexlab#1{#1}\fi
\providecommand{\url}[1]{\texttt{#1}}
\providecommand{\href}[2]{#2}
\providecommand{\path}[1]{#1}
\providecommand{\DOIprefix}{doi:}
\providecommand{\ArXivprefix}{arXiv:}
\providecommand{\URLprefix}{URL: }
\providecommand{\Pubmedprefix}{pmid:}
\providecommand{\doi}[1]{\href{http://dx.doi.org/#1}{\path{#1}}}
\providecommand{\Pubmed}[1]{\href{pmid:#1}{\path{#1}}}
\providecommand{\bibinfo}[2]{#2}
\ifx\xfnm\relax \def\xfnm[#1]{\unskip,\space#1}\fi
\bibitem[{Hirsimaki et~al.(2009)Hirsimaki, Pylkkonen, and Kurimo}]{bhref11}
\bibinfo{author}{T.~Hirsimaki}, \bibinfo{author}{J.~Pylkkonen},
  \bibinfo{author}{M.~Kurimo},
\newblock \bibinfo{title}{Importance of high-order n-gram models in morph-based
  speech recognition},
\newblock \bibinfo{journal}{IEEE Trans. Audio, Speech, and Language Processing}
  \bibinfo{volume}{17} (\bibinfo{year}{2009}) \bibinfo{pages}{724--732}.
\bibitem[{Varjokallio et~al.(2016)Varjokallio, Kurimo, and Virpioja}]{bh_ref10}
\bibinfo{author}{M.~Varjokallio}, \bibinfo{author}{M.~Kurimo},
  \bibinfo{author}{S.~Virpioja},
\newblock \bibinfo{title}{Class n-gram models for very large vocabulary speech
  recognition of finnish and estonian},
\newblock in: \bibinfo{booktitle}{Int. Conf. Stat. Lang. Speech Process.},
  \bibinfo{organization}{Springer}, \bibinfo{year}{2016}, pp.
  \bibinfo{pages}{133--144}.
\bibitem[{Enarvi and Kurimo(2016)}]{bhref12}
\bibinfo{author}{S.~Enarvi}, \bibinfo{author}{M.~Kurimo},
\newblock \bibinfo{title}{Theanolm-an extensible toolkit for neural network
  language modeling},
\newblock \bibinfo{journal}{arXiv preprint arXiv:1605.00942}
  (\bibinfo{year}{2016}).
\bibitem[{Hinton et~al.(2012)Hinton, Deng, Yu, Dahl, Mohamed, Jaitly, Senior,
  Vanhoucke, Nguyen, Sainath et~al.}]{hinton2012deep}
\bibinfo{author}{G.~Hinton}, \bibinfo{author}{L.~Deng},
  \bibinfo{author}{D.~Yu}, \bibinfo{author}{G.~E. Dahl}, \bibinfo{author}{A.-r.
  Mohamed}, \bibinfo{author}{N.~Jaitly}, \bibinfo{author}{A.~Senior},
  \bibinfo{author}{V.~Vanhoucke}, \bibinfo{author}{P.~Nguyen},
  \bibinfo{author}{T.~N. Sainath}, et~al.,
\newblock \bibinfo{title}{Deep neural networks for acoustic modeling in speech
  recognition: The shared views of four research groups},
\newblock \bibinfo{journal}{IEEE Sig. Process. Mag.} \bibinfo{volume}{29}
  (\bibinfo{year}{2012}) \bibinfo{pages}{82--97}.
\bibitem[{Hirsim{\"a}ki et~al.(2006)Hirsim{\"a}ki, Creutz, Siivola, Kurimo,
  Virpioja, and Pylkk{\"o}nen}]{bhref13}
\bibinfo{author}{T.~Hirsim{\"a}ki}, \bibinfo{author}{M.~Creutz},
  \bibinfo{author}{V.~Siivola}, \bibinfo{author}{M.~Kurimo},
  \bibinfo{author}{S.~Virpioja}, \bibinfo{author}{J.~Pylkk{\"o}nen},
\newblock \bibinfo{title}{Unlimited vocabulary speech recognition with morph
  language models applied to finnish},
\newblock \bibinfo{journal}{Computer Speech \& Language} \bibinfo{volume}{20}
  (\bibinfo{year}{2006}) \bibinfo{pages}{515--541}.
\bibitem[{Byrne et~al.(2000)Byrne, Haji{\v{c}}, Ircing, Krbec, and
  Psutka}]{bhref14}
\bibinfo{author}{W.~Byrne}, \bibinfo{author}{J.~Haji{\v{c}}},
  \bibinfo{author}{P.~Ircing}, \bibinfo{author}{P.~Krbec},
  \bibinfo{author}{J.~Psutka},
\newblock \bibinfo{title}{Morpheme based language models for speech recognition
  of czech},
\newblock in: \bibinfo{booktitle}{International Workshop on Text, Speech and
  Dialogue}, \bibinfo{organization}{Springer}, \bibinfo{year}{2000}, pp.
  \bibinfo{pages}{211--216}.
\bibitem[{Erdogan et~al.(2005)Erdogan, Buyuk, and Oflazer}]{bhref15}
\bibinfo{author}{H.~Erdogan}, \bibinfo{author}{O.~Buyuk},
  \bibinfo{author}{K.~Oflazer},
\newblock \bibinfo{title}{Incorporating language constraints in sub-word based
  speech recognition},
\newblock in: \bibinfo{booktitle}{IEEE Workshop on Automatic Speech Recognition
  and Understanding, 2005.}, \bibinfo{organization}{IEEE},
  \bibinfo{year}{2005}, pp. \bibinfo{pages}{98--103}.
\bibitem[{Laureys et~al.(2002)Laureys, Vandeghinste, and
  Duchateau}]{laureys2002hybrid}
\bibinfo{author}{T.~Laureys}, \bibinfo{author}{V.~Vandeghinste},
  \bibinfo{author}{J.~Duchateau},
\newblock \bibinfo{title}{A hybrid approach to compounds in lvcsr},
\newblock in: \bibinfo{booktitle}{VII Int. Conf. Spoken Language Processing},
  \bibinfo{year}{2002}.
\bibitem[{Hacioglu et~al.(2003)Hacioglu, Pellom, Ciloglu, Ozturk, Kurimo, and
  Creutz}]{hacioglu2003lexicon}
\bibinfo{author}{K.~Hacioglu}, \bibinfo{author}{B.~Pellom},
  \bibinfo{author}{T.~Ciloglu}, \bibinfo{author}{O.~Ozturk},
  \bibinfo{author}{M.~Kurimo}, \bibinfo{author}{M.~Creutz},
\newblock \bibinfo{title}{On lexicon creation for turkish lvcsr},
\newblock in: \bibinfo{booktitle}{Eighth European Conf. Speech Communication
  and Tech.}, \bibinfo{year}{2003}.
\bibitem[{Ar{\i}soy et~al.(2007)Ar{\i}soy, Sak, and
  Sara{\c{c}}lar}]{arisoy2007language}
\bibinfo{author}{E.~Ar{\i}soy}, \bibinfo{author}{H.~Sak},
  \bibinfo{author}{M.~Sara{\c{c}}lar},
\newblock \bibinfo{title}{Language modeling for automatic turkish broadcast
  news transcription},
\newblock in: \bibinfo{booktitle}{Eighth Annual Conf. Int. Speech Communication
  Assoc.}, \bibinfo{year}{2007}.
\bibitem[{Smit et~al.(2017)Smit, Virpioja, and Kurimo}]{sub_context}
\bibinfo{author}{P.~Smit}, \bibinfo{author}{S.~Virpioja},
  \bibinfo{author}{M.~Kurimo},
\newblock \bibinfo{title}{Improved subword modeling for {WFST}-based speech
  recognition},
\newblock in: \bibinfo{booktitle}{18th Ann. Conf. Int. Speech Communication
  Assoc. (INTERSPEECH 2017)}, \bibinfo{year}{2017}, pp.
  \bibinfo{pages}{2551--2555}.
\bibitem[{mic(2018)}]{microsoftdata}
\bibinfo{title}{Data provided by SpeechOcean.com and Microsoft},
  \bibinfo{publisher}{Microsoft}, \bibinfo{year}{2018}.
\bibitem[{Madhavaraj and Ramakrishnan(2017)}]{asr_tamil}
\bibinfo{author}{A.~Madhavaraj}, \bibinfo{author}{A.~G. Ramakrishnan},
\newblock \bibinfo{title}{Design and development of a large vocabulary,
  continuous speech recognition system for {Tamil}},
\newblock in: \bibinfo{booktitle}{2017 14th IEEE India Council International
  Conference (INDICON)}, \bibinfo{organization}{IEEE}, \bibinfo{year}{2017},
  pp. \bibinfo{pages}{1--5}.
\bibitem[{mil(2022{\natexlab{a}})}]{mile_Kannada_asr_data}
\bibinfo{title}{Openslr: Iisc mile kannada asr corpus},
  \bibinfo{year}{2022}{\natexlab{a}}. \URLprefix
  \url{http://www.openslr.org/126/}.
\bibitem[{mil(2022{\natexlab{b}})}]{mile_Tamil_asr_data}
\bibinfo{title}{{OpenSLR: IISc MILE Tamil ASR corpus}},
  \bibinfo{year}{2022}{\natexlab{b}}. \URLprefix
  \url{http://www.openslr.org/127/}.
\bibitem[{{Sarveswaran} et~al.(2018){Sarveswaran}, {Dias}, and
  {Butt}}]{sub_tamverbs}
\bibinfo{author}{K.~{Sarveswaran}}, \bibinfo{author}{G.~{Dias}},
  \bibinfo{author}{M.~{Butt}},
\newblock \bibinfo{title}{{ThamizhiFST: a morphological analyser and generator
  for Tamil verbs}},
\newblock in: \bibinfo{booktitle}{2018 3rd Int. Conf. Inf. Tech. Res. (ICITR)},
  \bibinfo{year}{2018}, pp. \bibinfo{pages}{1--6}.
\bibitem[{Agesthialingom(1971)}]{sub_wordformtam02}
\bibinfo{author}{S.~Agesthialingom},
\newblock \bibinfo{title}{A note on {Tamil} verbs},
\newblock in: \bibinfo{booktitle}{Anthropological Linguistics},
  volume~\bibinfo{volume}{13}, \bibinfo{year}{1971}, pp.
  \bibinfo{pages}{121--125}.
\bibitem[{Prathibha and Padma(2013)}]{prathibha2013development}
\bibinfo{author}{R.~Prathibha}, \bibinfo{author}{M.~Padma},
\newblock \bibinfo{title}{Development of morpholoical analyzer for kannada
  verbs}  (\bibinfo{year}{2013}).
\bibitem[{{Lushanthan} et~al.(2014){Lushanthan}, {Weerasinghe}, and
  {Herath}}]{sub_tamnouns}
\bibinfo{author}{S.~{Lushanthan}}, \bibinfo{author}{A.~R. {Weerasinghe}},
  \bibinfo{author}{D.~L. {Herath}},
\newblock \bibinfo{title}{Morphological analyzer and generator for {Tamil}
  language},
\newblock in: \bibinfo{booktitle}{2014 14th Int. Conf. Adv. {ICT} for Emerging
  Regions (ICTer)}, \bibinfo{year}{2014}, pp. \bibinfo{pages}{190--196}.
\bibitem[{Ramakrishnan and Narayana(2007)}]{ramakrishnan2007grapheme}
\bibinfo{author}{A.~Ramakrishnan}, \bibinfo{author}{M.~L. Narayana},
\newblock \bibinfo{title}{Grapheme to phoneme conversion for tamil speech
  synthesis},
\newblock in: \bibinfo{booktitle}{Proc. Workshop in Image and Signal Processing
  (WISP-2007), IIT Guwahati}, \bibinfo{year}{2007}, pp.
  \bibinfo{pages}{96--99}.

\end{thebibliography}
\end{document}